\documentclass[journal=jacsat,manuscript=article]{achemso}
\usepackage[version=3]{mhchem} 
\author{Sandip Aryal}
\email{saryal@lanl.gov}
\author{Enrique R. Batista}
\author{Gaoxue Wang}
\email{gaoxuew@lanl.gov}
\affiliation{Theoretical Division, Los Alamos National Laboratory, Los Alamos, New Mexico 87545, United States}
\title[An \textsf{achemso} demo]
  {Effect of Group-V Impurities on the Electronic Properties of  Germanium Detectors: An Insight from First-Principles Calculations}
\keywords{American Chemical Society, \LaTeX}
\begin{document}
%%%%%%%%%%%%%%%%%%%%%%%%%%%%%%%%%%%%%%%%%%%%%%%%%%%%%%%%%%%%%%%%%%%%%
%%%%%%%%%%%%%%%%%%%%%%%%%%%%%%%%%%%%%%%%%%%%%%%%%%%%%%%%%%%%%%%%%%%%%
%%%%%%%%%%%%%%%%%%%%%%%%%%%%%%%%%%%%%%%%%%%%%%%%%%%%%%%%%%%%%%%%%%%%%
%%%%%%%%%%%%%%%%%%%%%%%%%%%%%%%%%%%%%%%%%%%%%%%%%%%%%%%%%%%%%%%%%%%%%
\begin{abstract}
%%%%%%%%%%%%%%%%%%%%%%%%%%%%%%%%%%%%%%%%%%%%%%%%%%%%%%%%%%%%%%%%%%%%%
%%%%%%%%%%%%%%%%%%%%%%%%%%%%%%%%%%%%%%%%%%%%%%%%%%%%%%%%%%%%%%%%%%%%%
%%%%%%%%%%%%%%%%%%%%%%%%%%%%%%%%%%%%%%%%%%%%%%%%%%%%%%%%%%%%%%%%%%%%%

The outstanding properties of high-purity germanium (HPGe) detectors, such as excellent energy resolution, high energy sensitivity, and a low background-to-signal ratio, make them essential and ideal candidates for detecting particle signatures in nuclear processes such as neutrino-less double beta decay ($\textit{0}\nu\beta\beta$). However, the presence of defects and impurities in HPGe crystals can lead to charge trapping, which affects carrier mobility and results in significant energy resolution degradation.
In this work, we employ density functional theory with a hybrid functional to study the energetics of possible point defects in Ge. Our findings indicate that n-type group-V impurities, such as phosphorus (P), arsenic (As), and antimony (Sb), form more readily in Ge compared to nitrogen (N), Ge vacancies, and Ge interstitials. Unlike N dopants, which yield deep trap states, P, As, and Sb create shallow traps close to the conduction band edge of Ge. Furthermore, we predict that n-type defects can condense into defect complexes with Ge vacancies. These vacancy-impurity complexes form deep traps in Ge, similar to Ge vacancies, suggesting that both vacancies and vacancy-impurity complexes contribute to charge trapping in these detectors, thereby diminishing their performance.
%%%%%%%%%%%%%%%%%%%%%%%%%%%%%%%%%%%%%%%%%%%%%%%%%%%%%%%%%%%%%%%%%%%%%
%%%%%%%%%%%%%%%%%%%%%%%%%%%%%%%%%%%%%%%%%%%%%%%%%%%%%%%%%%%%%%%%%%%%%
%%%%%%%%%%%%%%%%%%%%%%%%%%%%%%%%%%%%%%%%%%%%%%%%%%%%%%%%%%%%%%%%%%%%%
\end{abstract}
%%%%%%%%%%%%%%%%%%%%%%%%%%%%%%%%%%%%%%%%%%%%%%%%%%%%%%%%%%%%%%%%%%%%%
%%%%%%%%%%%%%%%%%%%%%%%%%%%%%%%%%%%%%%%%%%%%%%%%%%%%%%%%%%%%%%%%%%%%%
%%%%%%%%%%%%%%%%%%%%%%%%%%%%%%%%%%%%%%%%%%%%%%%%%%%%%%%%%%%%%%%%%%%%%
%%%%%%%%%%%%%%%%%%%%%%%%%%%%%%%%%%%%%%%%%%%%%%%%%%%%%%%%%%%%%%%%%%%%%
%% Start the main part of the manuscript here
%%%%%%%%%%%%%%%%%%%%%%%%%%%%%%%%%%%%%%%%%%%%%%%%%%%%%%%%%%%%%%%%%%%%%
%%%%%%%%%%%%%%%%%%%%%%%%%%%%%%%%%%%%%%%%%%%%%%%%%%%%%%%%%%%%%%%%%%%%%
%%%%%%%%%%%%%%%%%%%%%%%%%%%%%%%%%%%%%%%%%%%%%%%%%%%%%%%%%%%%%%%%%%%%%
%%%%%%%%%%%%%%%%%%%%%%%%%%%%%%%%%%%%%%%%%%%%%%%%%%%%%%%%%%%%%%%%%%%%%
\section{I. Introduction}
%%%%%%%%%%%%%%%%%%%%%%%%%%%%%%%%%%%%%%%%%%%%%%%%%%%%%%%%%%%%%%%%%%%%%
%%%%%%%%%%%%%%%%%%%%%%%%%%%%%%%%%%%%%%%%%%%%%%%%%%%%%%%%%%%%%%%%%%%%%
High-purity Germanium (HPGe) detectors have proven to be the most sensitive among the current technologies for detecting the radiation and particle signatures of rare physical events in nature, such as neutrino-less double beta-decay ($\textit{0}\nu\beta\beta$). These detectors offer higher energy resolution, greater energy sensitivity, and lowest background signal among all detectors currently used in search of $\textit{0}\nu\beta\beta$.\cite{C01,C02} Observing $\textit{0}\nu\beta\beta$ would confirm the Majorana nature of neutrino (i.e. neutrino is its own antiparticle), leading to the violation of Lepton number \cite{C02,C03}. In that case, it would imply that neutrinos acquire mass through the effective Majorana neutrino mass term that is inversely proportional to $\textit{0}\nu\beta\beta$ half-lives. This is consistent with the observed discovery of neutrino oscillations in atmospheric and solar neutrinos oscillation experiments, which also indicate neutrinos to have non-zero mass.\cite{C02,C04, C05, C06, C07, C08} The widely used candidate for parent nuclei that can undergo $\textit{0}\nu\beta\beta$ is the Ge isotope $^{76}$Ge. \cite{C02, C08, C09, C10} HPGe detectors are built from high-purity Ge crystals isotopically enriched in $^{76}$Ge, serving both as sources and detectors of $\textit{0}\nu\beta\beta$.\cite{C11} These Ge crystals have a low impurity concentration  $<$ 10$^{10}$ cm$^{-3}$ and dislocation density $<$ 10$^4$ cm$^{-2}$.\cite{C12} They are used in several 
$\textit{0}\nu\beta\beta$ experiments such as GERDA \cite{C08, C09, C13, C14} and MAJORANA,\cite{C02,C03} and more recently in LEGEND\cite{C10,C15} collaboration, which combines efforts from the two. The LEGEND-200 collaboration aims to search for $\textit{0}\nu\beta\beta$ decay in enriched $^{76}$Ge HPGe detectors of mass 200 kg, targeting a $\textit{0}\nu\beta\beta$ decay half-life of 10$^{27}$ years.\cite{C06,C10} Likewise, LEGEND-1000 aims to develop ton-scale enriched $^{76}$Ge HPGe detector with a half-life of $\textit{0}\nu\beta\beta$ decay beyond 10$^{28}$ years.\cite{C10}

Despite the technical advantages of HPGe detectors, growing crystals free from defects and impurities is close to impossible. The incoming particles and radiations interacting with Ge atoms in these detectors create charge carriers that drift towards the electrodes. In large HPGe detectors, the long drift distance that these charge carriers migrate, increases the likelihood of them being trapped by impurities. Such charge trapping affects the performance of these detectors by degrading their energy resolution.\cite{C16} Therefore, understanding the role of defects and impurities in crystalline Ge is crucial for optimizing the fabrication process of these detectors and improving their energy resolution by eliminating defects during manufacturing. Although a low concentration of defects is desirable for optimal performance in HPGe detectors, contamination with group-V impurities such as nitrogen (N), phosphorus (P), arsenic (As), and antimony (Sb) can occur during the manufacturing process. Furthermore, previous studies have reported that n-type impurities, along with Ge vacancies, are the most common defects in Ge. \cite{C17, C18, C19, C20, C21, A57, A58, A59, A52, A53, A54, A55, A56}

In this work, we employ first-principles calculations using a hybrid functional within density functional theory (DFT) to study the electronic structures of possible n-type defects and their complexes in Ge. A comprehensive assessment of the energetics of the point defects considered in this study reveals that P, As, and Sb are the most favorable defects in Ge compared to N, Ge vacancies, and Ge interstitials. Unlike N, Ge vacancies, and Ge interstitials, which have charge transition levels far from the band edges, forming deep traps, P, As, and Sb introduce ionization levels close to the conduction band edge, resulting in shallow traps in Ge. Since these n-type impurities diffuse in Ge through their interaction with vacancies,\cite{C18,C19, C20, C21, A53, A55, A56} we also investigated the formation of defect complexes in Ge.
We infer from our results that the n-type impurities (P, As, Sb) can form complexes with vacancies; larger complexes are found to bind strongly than the smaller ones. Furthermore, such clustering lowers their formation energy, as larger complexes of n-type impurities (P, As, Sb) with Ge vacancies have lower formation energies than smaller ones. The impurity-vacancy complexes introduce charge transition levels far from the band edges, forming deep traps in Ge. Consequently, these defect complexes are expected to act as charge-trapping centers, potentially degrading the performance of HPGe detectors and their ability to detect particle signatures in rare physical processes in nature.

%%%%%%%%%%%%%%%%%%%%%%%%%%%%%%%%%%%%%%%%%%%%%%%%%%%%%%%%%%%%%%%%%%%%%
%%%%%%%%%%%%%%%%%%%%%%%%%%%%%%%%%%%%%%%%%%%%%%%%%%%%%%%%%%%%%%%%%%%%%
%%%%%%%%%%%%%%%%%%%%%%%%%%%%%%%%%%%%%%%%%%%%%%%%%%%%%%%%%%%%%%%%%%%%%
\section{II. Computational Details}
%%%%%%%%%%%%%%%%%%%%%%%%%%%%%%%%%%%%%%%%%%%%%%%%%%%%%%%%%%%%%%%%%%%%%
%%%%%%%%%%%%%%%%%%%%%%%%%%%%%%%%%%%%%%%%%%%%%%%%%%%%%%%%%%%%%%%%%%%%%
%%%%%%%%%%%%%%%%%%%%%%%%%%%%%%%%%%%%%%%%%%%%%%%%%%%%%%%%%%%%%%%%%%%%%
\subsection{A. First-principles calculations}
We employed spin-polarized plane-wave DFT as implemented in VASP (version 6.1.2),\cite{C22, C23} in conjunction with the open-source python package SPINNEY,\cite{C24} to study charged defects in Ge. The local density approximation (LDA) was used to approximate the exchange-correlation energy functional during structural relaxations. The projector augmented wave (PAW) method\cite{C28, C29} was employed to account for the interactions between valence electrons and ionic cores. 
To model point defects—vacancies, interstitials, and substitutional n-type dopants—we constructed a 3×3×3 supercell of Ge containing 216 atoms. Structural relaxations were performed at the LDA level without any symmetry constraints until the residual force on each atom was below 0.001 eV/{\AA}.
Since the local and semi-local DFT functionals  are known to underestimate the bandgap of Ge due to self-interaction errors associated with these functionals, we employed a single point hybrid functional (HSE06)\cite{C25, C26} calculations on LDA relaxed structures to study the energetics of these defects. 

The HSE06 functional, which partially corrects the self-interaction errors associated with the standard DFT functionals by incorporating a fraction ($\sim$25\%) of Hartree-Fock exchange term, is known to yield accurate electronic bandgaps and lattice constants for semiconductors.\cite{C25, C26} It is noteworthy that HSE06 lattice constant of 5.61 {\AA} for Ge agrees very well with LDA lattice constant of 5.65 {\AA}, within 1\%, suggesting that the structural properties predicted by LDA are comparable to those from HSE06 for Ge. This further justifies the use of LDA relaxed structures for single-point HSE06 calculations of these charged defects. A Monkhrost-Pack\cite{C27} k-point mesh of 2x2x2 and a plane-wave basis set with energy cutoff of 350 eV were used in these calculations. The  total energy convergence criterion during self-consistent cycle was set to $\rm 10^{-6}$ eV. 

\subsection{B. Formalism for defect calculations}
The formation energy of a charged defect was computed using \cite{C24, C30, A47, A48}
\begin{equation}
 \rm \Delta{E_f} = E_{D,q} - E_H - \sum_i{n_i}\mu_i
 + q(\epsilon_{v} + E_F) + E_{corr}
 \label{E1}
\end{equation}
where $\rm E_{D,q}$ is the total energy of a supercell containing defect, D, in charge state q; $\rm E_H$ is the energy of the pristine supercell; and $\rm n_i$ is the number of atoms of type i added ($\rm n_i > 0$) or removed ($\rm n_i < 0$) in the process of forming the defect. The chemical potential of species i, $\rm \mu_i = \mu_i^0 + \Delta\mu_i$, describes the exchange of particles with respective reservoirs and depends on the reference chemical potential of atomic species i, $\rm \mu_i^0$. Here, $\rm \Delta\mu_i$ is any deviation from the reference state. The reference chemical potential can be computed from respective standard elemental phase using DFT. For Ge, bulk Ge serves as the standard phase, yielding $\rm \mu_{Ge}^0$ = -5.68 eV/atom. Likewise, for N, the molecular gas phase is used as reference, giving $\rm \mu_{N}^0$ = -10.49 eV/atom. For P, As, and Sb, the reference is their respective crystalline phase. Our calculations give $\rm \mu_{P}^0$ =  -6.54 eV/atom, $\rm \mu_{As}^0$ = -5.95 eV/atom, and $\rm \mu_{Sb}^0$  = -5.40 eV/atom, all obtained using HSE06 functional. 

In Equation \ref{E1}, $\rm \epsilon_v$ is the valence band maximum, and $\rm E_F$  is electron chemical potential. $\rm E_F$ is confined within the bandgap of Ge, such that 0 $\leq$ $\rm E_F$ $\leq$ $\rm E_g$, where 0 represents the position of the valence band maximum and $\rm E_g$ represents the bandgap of Ge. The position of $\rm E_F$ in semiconductors and insulators depends on the level of doping and presence of external fields. $\rm E_{corr}$, in Equation \ref{E1}, is the correction term that takes into account of finite-size effects arising due to the use of supercells in our calculations. We used the scheme of Kumagai and Oba,\cite{C30} as implemented in the python package SPINNEY,\cite{C24} to compute the finite-size corrections. The correction term can be expressed as: 
\begin{equation}
    \rm E_{corr} = -E_{lat} + q\Delta\phi,
\end{equation}
 where $\rm E_{lat}$ is the Madelung energy of charge density $\rho_d$ embedded in the host material in presence of jellium background and $\Delta\phi$ is the potential alignment term.\cite{C24}

As a function of $\rm E_F$, Equation \ref{E1} describes a straight line with a slope corresponding to the defect's charge state q. The diagram showing the variation of formation energy of a charged defect as a function of $\rm E_F$ for different q is the defect diagram or charge transition level diagram. The primary purpose of this diagram is to identify charge transition levels or ionization levels, which correspond to the Fermi level positions where the formation energies of two charge states become equal.
The charge transition level, $\rm \epsilon{(q/q^{\prime})}$, is expressed in terms of the formation energies of defects in charge states q and $\rm {q^{\prime}}$  as\cite{A48}
\begin{equation}
 \rm  \epsilon(q/q^{\prime}) = \frac{\Delta{E_f} (D,q; E_F=0) - \Delta{E_f}(D,{q^{\prime}}; E_F=0)}{q^{\prime} - q}
 \label{E2}
\end{equation}
where, $\rm \Delta{E_f}(D,q, E_F=0)$  and $\rm \Delta{E_f}(D,{q^{\prime}}, E_F=0)$ are formation energies for defects in charge state q and $\rm q^{\prime}$ at the valence band maximum ($\rm E_F$ = 0).
A charge transition level positioned near the band edge signifies a shallow trap. Typically, shallow traps are within a few $\rm k_B$T ($\leq$ 0.1 eV) from the band edges, where $\rm k_B$ is the Boltzmann constant and T is the absolute temperature. Conversely, when an ionization level is significantly far from the band edges, it indicates a deep trap.  In general, deep traps hinder the charge transport and degrade optical properties by facilitating charge trapping and boosting non-radiative recombination in semiconductors and insulators.
%%%%%%%%%%%%%%%%%%%%%%%%%%%%%%%%%%%%%%%%%%%%%%%%%%%%%%%%%%%%%%%%%%%%%
%%%%%%%%%%%%%%%%%%%%%%%%%%%%%%%%%%%%%%%%%%%%%%%%%%%%%%%%%%%%%%%%%%%%%
%%%%%%%%%%%%%%%%%%%%%%%%%%%%%%%%%%%%%%%%%%%%%%%%%%%%%%%%%%%%%%%%%%%%%
\section{III. Results and Discussion}
%%%%%%%%%%%%%%%%%%%%%%%%%%%%%%%%%%%%%%%%%%%%%%%%%%%%%%%%%%%%%%%%%%%%%
%%%%%%%%%%%%%%%%%%%%%%%%%%%%%%%%%%%%%%%%%%%%%%%%%%%%%%%%%%%%%%%%%%%%%
%%%%%%%%%%%%%%%%%%%%%%%%%%%%%%%%%%%%%%%%%%%%%%%%%%%%%%%%%%%%%%%%%%%%%
\subsection{A. Crystalline Ge}

We begin with studying the structural and electronic properties of crystalline Ge. Ge adopts diamond-like structure with an FCC lattice and space group $\rm Fd\bar{3}$m (No. 227). 
The primitive basis in this crystal structure consists of two identical Ge atoms located at (0,0,0) and (1/4, 1/4, 1/4), associated with each FCC lattice point.\cite{A51} The primitive and conventional unit cell structures of Ge are shown in Fig. S1. The computed lattice parameter for Ge at LDA level in our calculations is 5.65 {\AA}, which agrees well with the experimental lattice constant of 5.66 {\AA} \cite{C31} within 0.2 \%. The Ge-Ge bond length in our calculations is 2.45 {\AA}. However, LDA predicts a semi-metallic behavior for Ge due to self-interaction errors associated with pure DFT local and semi-local functionals.  The hybrid functional HSE06, which corrects these inaccuracies, yields an indirect bandgap of 0.79 eV for Ge (Fig. S1) at the HSE06 relaxed lattice constant $\sim$5.61 {\AA}. The computed bandgap in this work is in close agreement with the experimental bandgap of 0.740-0.785 eV at 0 K.\cite{C20, C32, C33} 
%%%%%%%%%%%%%%%%%%%%%%%%%%%%%%%%%%%%%%%%%%%%%%%%%%%%%%%%%%%%%%%%%%%%%%%
%%%%%%%%%%%%%%%%%%%%%%%%%%%%%%%%%%%%%%%%%%%%%%%%%%%%%%%%%%%%%%%%%%%%%%%
%%%%%%%%%%%%%%%%%%%%%%%%%%%%%%%%%%%%%%%%%%%%%%%%%%%%%%%%%%%%%%%%%%%%%%%
%%%%%%%%%%%%%%%%%%%%%%%%%%%%%%%%%%%%%%%%%%%%%%%%%%%%%%%%%%%%%%%%%%%%%%%
%%%%%%%%%%%%%%%%%%%%%%%%%%%%%%%%%%%%%%%%%%%%%%%%%%%%%%%%%%%%%%%%%%%%%%%
%%%%%%%%%%%%%%%%%%%%%%%%%%%%%%%%%%%%%%%%%%%%%%%%%%%%%%%%%%%%%%%%%%%%%%%
\subsection{B. Ge vacancy, Ge interstitial, and n-type substitutional dopants}
%%%%%%%%%%%%%%%%%%%%%%%%%%%%%%%%%%%%%%%%%%%%%%%%%%%%%%%%%%%%%%%%%%%%%%%
%%%%%%%%%%%%%%%%%%%%%%%%%%%%%%%%%%%%%%%%%%%%%%%%%%%%%%%%%%%%%%%%%%%%%%%
%%%%%%%%%%%%%%%%%%%%%%%%%%%%%%%%%%%%%%%%%%%%%%%%%%%%%%%%%%%%%%%%%%%%%%%
%%%%%%%%%%%%%%%%%%%%%%%%%%%%%%%%%%%%%%%%%%%%%%%%%%%%%%%%%%%%%%%%%%%%%%%
%%%%%%%%%%%%%%%%%%%%%%%%%%%%%%%%%%%%%%%%%%%%%%%%%%%%%%%%%%%%%%%%%%%%%%%
%%%%%%%%%%%%%%%%%%%%%%%%%%%%%%%%%%%%%%%%%%%%%%%%%%%%%%%%%%%%%%%%%%%%%%%
\subsection{Structural Properties}
\begin{figure}[ht!]
\centering
\includegraphics[width=8cm]{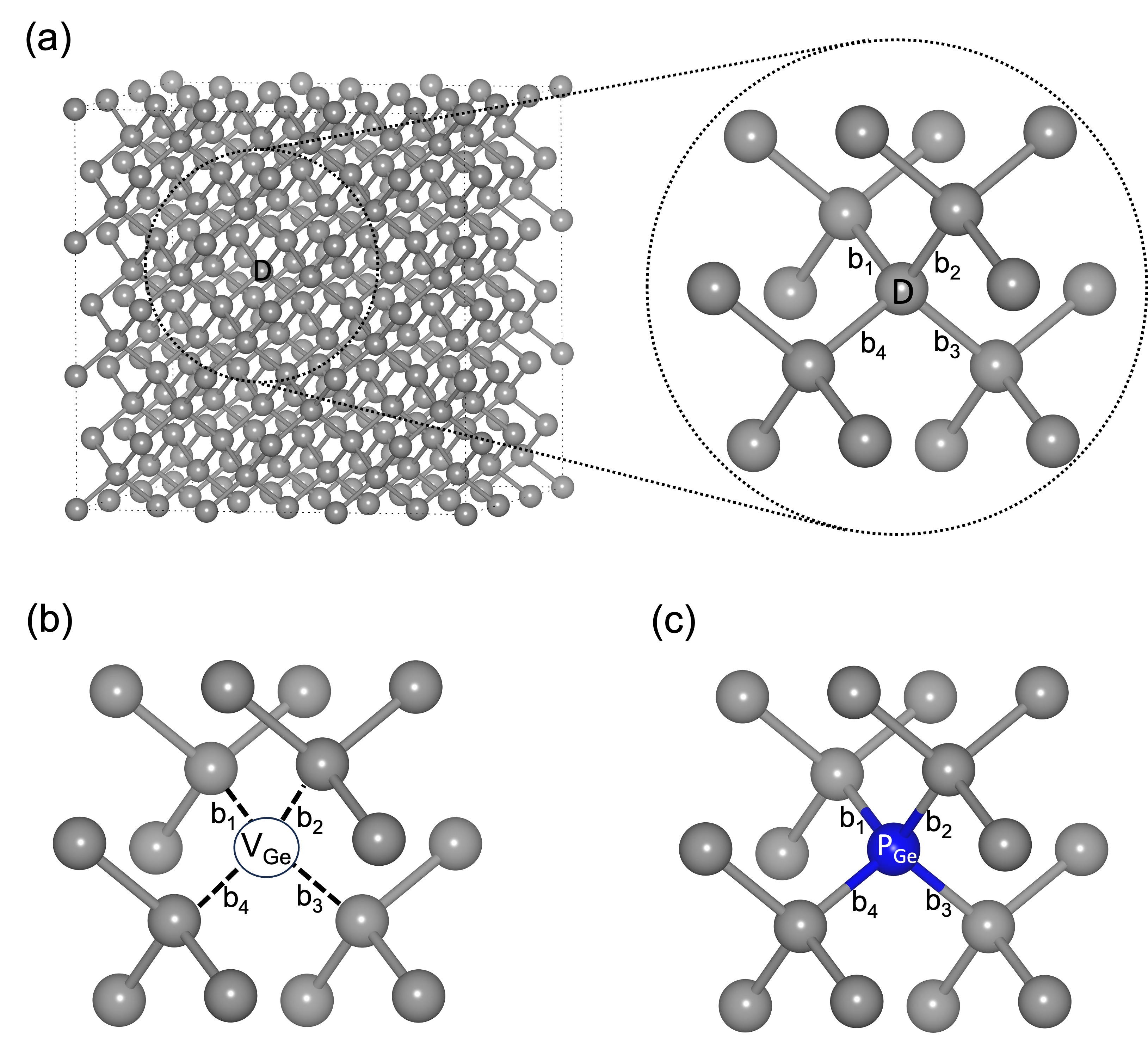}
\caption{(a) The 216 atoms supercell of Ge constructed from its conventional unit cell. The supercell is zoomed in near the defect site, D, to provide a clear view of the nearest Ge atoms bonded to D. In the zoomed figure, $\rm b_1$, $\rm b_2$, $\rm b_3$, and $\rm b_4$ show the bond length of the Ge atom at site D with its nearest neighbors. For a pristine supercell of Ge, $\rm b_1$=$\rm b_2$=$\rm b_3$=$\rm b_4$=2.45 {\AA},  (b) Ge vacancy ($\rm V_{Ge}$), and (c) P substitutional defect at D ($\rm P_{Ge}$). The gray and blue spheres represent Ge and P atoms, respectively. }
\label{fig:Fig1}
\end{figure}

The simulations of point defects including Ge vacancy, Ge interstitials, substitutional n-type dopants, and their complexes, were carried out in a large 3x3x3 supercell (Fig. \ref{fig:Fig1}(a)) consisting of 216 Ge atoms. This supercell was constructed from the conventional unit cell of Ge using the LDA-relaxed lattice constant ($a$ $\sim$5.65 {\AA}). Such large supercell ($a$ = $\sim$16.95 {\AA}) minimizes interactions between a defect and its periodic images when employing the periodic supercell approach in our calculations. In Fig. \ref{fig:Fig1}(a), the supercell is zoomed in near the defect site (D), to clearly visualize the adjacent Ge atoms bonded to D. A vacancy in this supercell is created by removing a Ge atom at site D, as shown in Fig. \ref{fig:Fig1}(b) and is labeled by $\rm V_{Ge}$. Fig. \ref{fig:Fig1}(c) illustrates a single Ge atom substituted by a phosphorus (P) atom, represented by $\rm P_{Ge}$. Similarly, other n-type defects like Nitrogen (N), Arsenic (As), and Antimony (Sb) are labeled by $\rm N_{Ge}$, $\rm As_{Ge}$, and $\rm Sb_{Ge}$. The defective supercells were then relaxed at LDA level of theory, which, as confirmed above, yields similar structures  comparable to those obtained using HSE06. Upon relaxation, these defect supercells undergo structural changes near defect sites; however, in all cases, we found these relaxations to be short-range.

%%%%%%%%%%%%%%%%%%%%%%%%%%%%%%%%%%%%%%%%%%%%%%%%%%%%%%%%%%%%%%%%%%%%%%%
\begin{table}[ht!]
\caption{\label{tab:table1} Calculated defect-Ge bond lengths at LDA level. Our calculations give $\rm b_1$=$\rm b_2$=$\rm b_3$=$\rm b_4$ (see Figure \ref{fig:Fig1}).}
\centering
\begin{tabular}{c c c c c c c} % centered columns (4 columns)
\hline\hline
Defect Type   & $\rm  V_{Ge}  $ & $\rm  P_{Ge}$ & $\rm N_{Ge}$ & $\rm As_{Ge}$ & $\rm Sb_{Ge}$ & Crystalline Ge\\
\hline
Bond Type   & $\rm  V$-$\rm Ge  $ & $\rm  P$-$\rm Ge$ & $\rm N$-$\rm Ge$ & $\rm As$-$\rm Ge$ & $\rm Sb$-$\rm Ge$ & Ge$-$Ge\\
\hline
 Bond length  & \rm 1.94 {\AA}  & 2.41 {\AA}  & \rm 2.11 {\AA}  & 2.49 {\AA}  & 2.62 {\AA}  & 2.45 {\AA}  \\
\hline\hline
\end{tabular}
\end{table}
%%%%%%%%%%%%%%%%%%%%%%%%%%%%%%%%%%%%%%%%%%%%%%%%%%%%%%%%%%%%%%%%%%%%%%%

Depending on the type of defect (vacancy or substitutional) and the radius of the substituted atom, the bond distances between the defect  and the nearest Ge atoms change. Table \ref{tab:table1} summarizes the calculated defect-Ge bond lengths obtained in this work. As seen in Table \ref{tab:table1}, $\rm V_{Ge}$, $\rm P_{Ge}$ and $\rm N_{Ge}$ result in  bond length contraction in the vicinity of the defect site D, while the other n-type defects, $\rm As_{Ge}$ and $\rm Sb_{Ge}$, cause the bond length elongation. The variations in defect-Ge bond lengths   for different n-type defects (see Table \ref{fig:Fig1}) originate from the difference in their atomic radii and electronegativity. The smaller size of N and P atoms, combined with their larger electronegativity difference relative to Ge, leads to a contraction of the N-Ge and P-Ge bond lengths. Conversely, the larger atomic size of As and Sb, along with their smaller electronegativity difference with Ge, results in an elongation of the As-Ge and Sb-Ge bond lengths. Far from the defect site, Ge-Ge bond length remains unchanged,  indicating that these defects only affect the local structural properties of crystalline Ge.

\subsection{Formation energy and charge transition level for a Ge vacancy}
%We start with calibrating our calculations by comparing the obtained charge transition levels for a vacancy in Ge ($\rm V_{Ge}$) with other works. 
Previous theoretical studies\cite{C35,C36,C37,C38} have investigated charged vacancies in Ge ($\rm V_{Ge}$) using both DFT and beyond DFT methods. Table \ref{tab:table0} presents the calculated formation energy of  a charged $\rm V_{Ge}$ at valence band maximum ($\rm E_F = 0$). The computed formation energy is in reasonable agreement with previous works,  considering that those works employed  different levels of approximations and supercell sizes. 
Notably, when comparing the formation energy of $\rm V_{Ge}^{0}$ with reported value in Ref. \citenum{A50}, we find excellent agreement  (see Table \ref{tab:table0}). Our calculations indicate that a supercell containing at least 216 atoms is required to achieve convergence in the formation energy of a $\rm V_{Ge}$. Therefore, in this work, all defect calculations were performed using a 216-atom supercell of Ge.

%%%%%%%%%%%%%%%%%%%%%%%%%%%%%%%%%%%%%%%%%%%%%%%%%%%%%%%%%%%%%%%%%%%%%%%
\begin{table}[ht!]
\caption{\label{tab:table0} Comparison of computed formation energy of a charged $\rm V_{Ge}$ with other works. The calculations were performed at $\rm E_F =0$.}
\centering
\begin{tabular}{c c c c c c c c } % centered columns (4 columns)
\hline\hline
Method & $N_{atoms}$ & $\rm V_{Ge}^{-2}$  &  $\rm V_{Ge}^{-1}$ & $\rm V_{Ge}^{0}$ & $\rm V_{Ge}^{+1}$ & $\rm V_{Ge}^{+2}$ \\
\hline
LDA\cite{C36} & 64 & 2.53  & 2.28  & 2.28  & 2.62 & 3.36     \\ %PHYSICAL REVIEW B VOLUME 
HSE\cite{C37}& 64 & 4.03  & 3.38  & 2.87  & 3.34 & 3.98     \\ %PHYSICAL REVIEW B VOLUME 
HSE\cite{A50} & 216 & -  & -  & 3.70  & - & -     \\ %PRL 108, 066404 (2012) 
This work (HSE)& 64  & -  & -  & 3.48 & - & -  \\
This work (HSE)& 512  & -  & -  & 3.74 & - & - \\
This work (HSE)& 216  & 4.40  & 4.06  & 3.82 & 4.02 & 4.30  \\
\\ %PHYSICAL REVIEW B VOLUME 
\hline\hline
\end{tabular}
\end{table}
%%%%%%%%%%%%%%%%%%%%%%%%%%%%%%%%%%%%%%%%%%%%%%%%%%%%%%%%%%%%%%%%%%%%%%%

\begin{figure}[ht!]
\centering
\includegraphics[width=10cm]{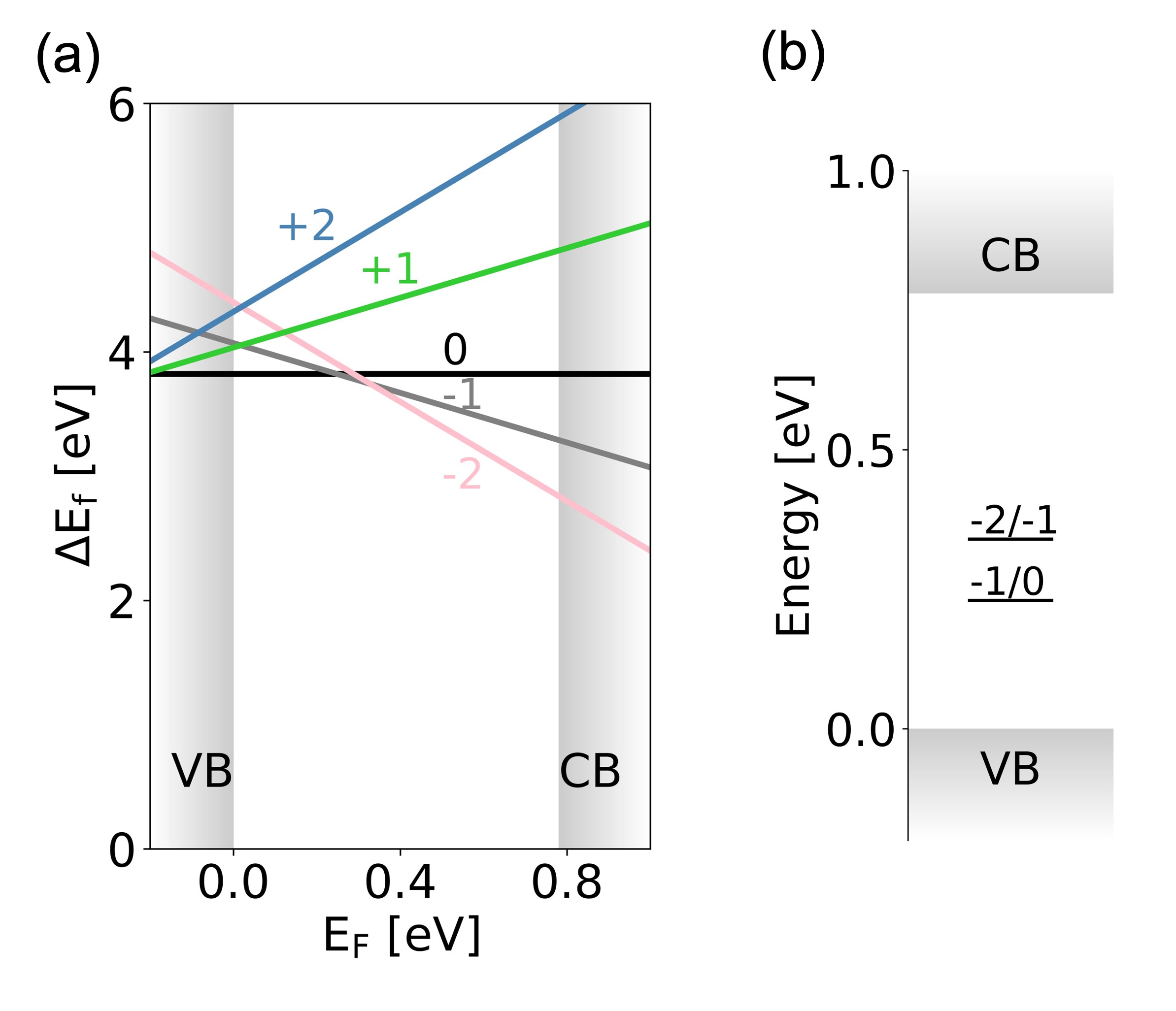}
\caption{(a) Formation energy ($\rm \Delta E_f$) as a function of the Fermi energy ($\rm E_F$) for a $\rm V_{Ge}$ (b) Charge transition levels for a $\rm V_{Ge}$. The shaded region in plots represents the valence band (VB) and conduction band (CB).}
\label{fig:Fig0}
\end{figure}

Fig. \ref{fig:Fig0}(a) illustrates the formation energy of $\rm V_{Ge}$ as a function of electron chemical potential for various charge states: -2, -1, 0, +1, and +2. Our findings indicate that the +2 and +1 charge states of $\rm V_{Ge}$ exhibit higher formation energies compared to the -1, 0, and -2 charge states.
Therefore, $\rm V_{Ge}$ if exists, is most likely to be in -1, 0, and -2 charge states, depending upon the position of the fermi energy within the bandgap of Ge. Our calculations further reveal that $\rm V_{Ge}^{-2}$ is the dominant defect for n-type doping conditions, aligning with previous theoretical \cite{C20} and experimental \cite{C17} studies.
Under the p-type doping conditions,  $\rm V_{Ge}^{0}$ emerges as the most stable defect, consistent with earlier theoretical research.\cite{C20}

%%%%%%%%%%%%%%%%%%%%%%%%%%%%%%%%%%%%%%%%%%%%%%%%%%%%%%%%%%%%%%%%%%%%%%%
Fig. \ref{fig:Fig0} also depicts that $\rm V_{Ge}$ behaves as a multi-level acceptor. The computed charge transition levels are 0.23 eV for the -1/0 transition and 0.34 eV for the -2/-1 transition, suggesting that $\rm V_{Ge}$ introduces deep traps in Ge. Such deep traps are detrimental as they can lead to charge trapping, ultimately degrading the energy resolution of HPGe detectors. The obtained charged transition levels are shown in Fig. \ref{fig:Fig0}(b).

Table \ref{tab:table0} compares the charge transition levels of a $\rm V_{Ge}$ obtained in this work with previous theoretical and experimental findings. 
The -1/0 charge transition level (0.23 eV) predicted in this work closely aligns with corresponding experimental value of 0.20 $\pm$ 0.04 eV, measured with the Perturbed Angular Correlation Spectroscopy (PAC),\cite{C39} demonstrating the reliability of our results.
Similarly, another experimental study utilizing Deep Level Transient Spectroscopy (DLTS) identified a transition level at 0.33 eV\cite{C40} above the valence band maximum for a $\rm V_{Ge}$. Comparing our theoretical predictions with the DLTS data, we infer that the ionization level at 0.33 eV likely corresponds to -2/-1 transition for a $\rm V_{Ge}$. 

%%%%%%%%%%%%%%%%%%%%%%%%%%%%%%%%%%%%%%%%%%%%%%%%%%%%%%%%%%%%%%%%%%%%%%%
\begin{table}[ht!]
\caption{\label{tab:table2} Comparison of charge transition level (eV) for a Ge vacancy with previous works.}
\centering
\begin{tabular}{c c c c c c c } % centered columns (4 columns)
\hline\hline
Method  & $N_{atoms}$  &  +1/+2 & 0/+1 & $ $-$1/0$ & $ $-$2/$-$1 $ \\
\hline
LDA\cite{C35} & 128  & -  & 0.21  & 0.37 & 0.40     \\ %PHYSICAL REVIEW B VOLUME 61, NUMBER 4 15 JANUARY 2000-II Fazio et al.
%LDA+U\cite{C36}  & 64 & -  & -  & 0.02  & 0.26  \\ %JOURNAL OF APPLIED PHYSICS 103, 08610320 08
GGA+U\cite{C20}   & 64  & -  & - & 0.21 & 0.27     \\ %Appl. Phys. Lett. 99, 072112 (2011)
HSE\cite{C37}  & 64 & -  & -  & 0.50  & 0.65   \\ %J. Appl. Phys. 110, 063534 (2011)
HSE\cite{C38}  & 216 & 0.14 & 0.15 & 0.16  & 0.38    \\ %PHYSICAL REVIEW B 87, 035203 (2013)
HSE\cite{A50}  & 216 & - & - & -  & 0.33   \\ %PHYSICAL REVIEW B 87, 035203 (2012)
This work (HSE)& 216 & - & -  & 0.23  & 0.34    \\
Experiment\cite{C39}  & - & - & -  & $\rm 0.20\pm0.04$  & -    \\
\hline\hline
\end{tabular}
\end{table}
%%%%%%%%%%%%%%%%%%%%%%%%%%%%%%%%%%%%%%%%%%%%%%%%%%%%%%%%%%%%%%%%%%%%%%%

The computed -2/-1 transition for a $\rm V_{Ge}$ in this study is in excellent agreement with the value reported in Ref. \citenum{A50}, which was obtained using the HSE functional with a similar supercell size and k-grid sampling.
 However, Ref. \citenum{A50} does not report the -1/0 charge transition level. 
Another study (Ref. \citenum{C38}) investigated the charged $\rm V_{Ge}$ at the HSE level using a 216-atom Ge supercell, while the calculations are done at $\Gamma$-point. Our computed -1/0 and -2/-1 charge transition levels are in close agreement (within 0.05 eV) with those reported in Ref. \citenum{C38}. However, Ref. \citenum{C38} also identifies additional 0/+1 and +1/+2 charge transition levels at 0.15 eV and 0.14 eV, respectively, which are not observed in our calculations.
This discrepancy is likely due to the exclusive use of the $\Gamma$-point in their simulations.  Our calculations indicate that a 2×2×2 k-grid is required to achieve convergence in the formation energy and charge transition levels for a 3×3×3 Ge supercell. 
Additionally, another study (Ref. \citenum{C37}) examined charged $\rm V_{Ge}$  at the HSE level using a smaller 64-atoms Ge supercell and reported -1/0 and -2/-1 transition levels at 0.50 eV and 0.65 eV, respectively. The deviation between our results and those in Ref. \citenum{C37} can be attributed to the smaller supercell used in that study. As previously noted, a supercell with at least 216 atoms is necessary to achieve convergence in the formation energy and ionization levels of defects in Ge.

Another study conducted at LDA level (Ref. \citenum{C35}) reports the -1/0 and -2/-1 charge transition levels at 0.37 eV and 0.40 eV, respectively. In addition, they identify a 0/+1 transition at 0.21 eV, which is not observed in our calculations. The small energy difference between the -1/0 and -2/-1 transitions in their study is a result of ignoring spin-polarization in their calculations. The spacing between the charge transition levels is associated with intra-orbital repulsion, and so the spin-polarization is essential to get the correct spacing between them.\cite{C38}
Other factors contributing to the discrepancies between their results and ours include the use of the LDA functional, a smaller 128-atom supercell, and simulations performed exclusively at the $\Gamma$-point.
The -1/0 transition level obtained using the GGA+U method in Ref. \citenum{C20} is in reasonable agreement with our results, despite their use of a 64-atom Ge supercell. 
 However, the obtained -2/-1 transition in Ref. \citenum{C20} lies close to the -1/0 transition level compared to this work. 

\subsection{Formation energy and charge transition level for Group-V defects}
Since we validated our approach by comparing our results for a $\rm V_{Ge}$ with previous experimental and theoretical studies, we now proceed to investigate the energetics of potential n-type defects in Ge.
\begin{figure}[ht!]
\centering
\includegraphics[width=14cm]{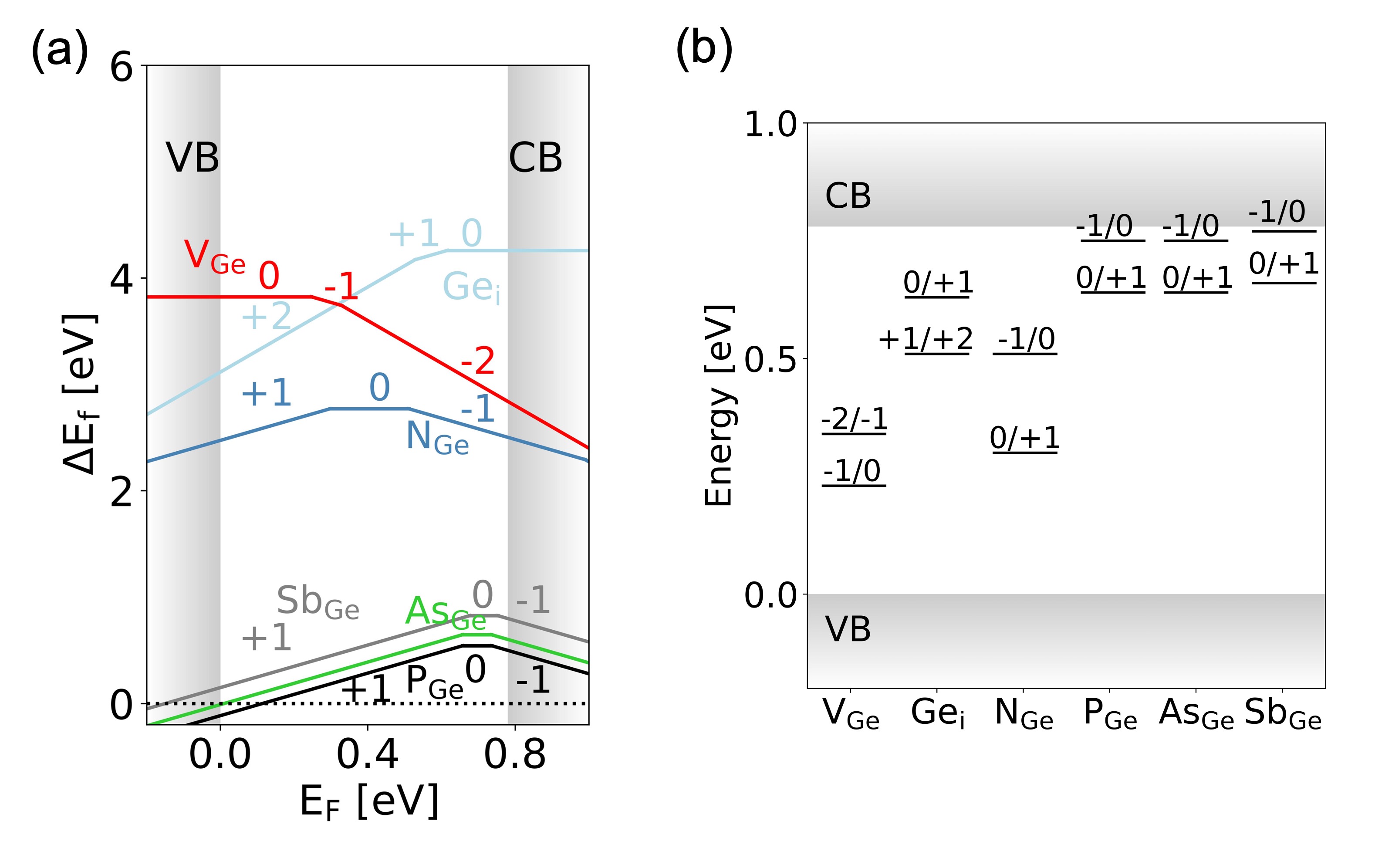}
\caption{(a) Formation energy ($\rm \Delta E_f$) as a function of the Fermi energy ($\rm E_F$) for n-type group-V dopants in Ge (b) Charge transition levels for n-type dopants in Ge. For comparison purposes, results for $\rm V_{Ge}$ and $\rm Ge_{i}$ are included in the plots. The shaded region in plots represents the valence band (VB) and conduction band (CB).}
\label{fig:Fig2}
\end{figure}
Group-V n-type dopants are among the most common defects in Ge crystals \cite{C17, C18, C19, C20, C21, A57, A58, A59, A52, A53, A54, A55, A56} 
  and are possible contaminants in HPGe detectors during the detector manufacturing process. 

Fig. \ref{fig:Fig2}(a) presents the formation energy as a function of electron chemical potential for different substitutional n-type dopants in Ge. For comparison, we also included our results for $\rm V_{Ge}$ and $\rm Ge_{i}$ at tetrahedral site in Fig. \ref{fig:Fig2}(a).

We infer from Fig. \ref{fig:Fig2}(a) that the formation energy of n-type defects in Ge is significantly lower than that of $\rm V_{Ge}$ and $\rm Ge_{i}$. Among the n-type impurities examined in this work,  \(\rm P_{Ge}\) exhibits a much lower formation energy than \(\rm N_{Ge}\) and is comparable in magnitude to \(\rm As_{Ge}\) and \(\rm Sb_{Ge}\). The formation energy of a defect is directly related to its equilibrium concentration ($\rm c_q^D$) as: $\rm c_q^D$ = $\rm N_{sites}$$\rm exp(-\Delta E_{f}(D,q)/{k_BT}$\cite{A48,A49}, where  $\rm N_{sites}$ is the appropriate site concentration for the defect. Since the formation energy of $\rm P_{Ge}$ is significantly lower than that of $\rm N_{Ge}$, $\rm V_{Ge}$, and $\rm Ge_{i}$, its concentration in Ge  is expected to be much higher. Due to very similar formation energies, $\rm As_{Ge}$ and $\rm Sb_{Ge}$ will have similar concentration as $\rm P_{Ge}$, indicating that these defects (\(\rm P_{Ge}\), \(\rm As_{Ge}\), and \(\rm Sb_{Ge}\)) are the most prevalent defects in Ge among those considered in this study. Fig. \ref{fig:Fig2}(a) also depicts that the n-type defects considered here exhibit +1/0 and -1/0 charge transitions within the bandgap of Ge; their energetic locations depend on the type of dopant. The charge transition levels for \(\rm P_{Ge}\), \(\rm As_{Ge}\), and \(\rm Sb_{Ge}\) are located near the conduction band edge, approximately at 0.64 eV and 0.75 eV for the respective 0/+1 and -1/0 transitions. $\rm N_{Ge}$, however, exhibits  0/+1 charge transition at $\sim$0.30 eV and -1/0 transition at $\sim$0.51 eV. The ionization levels of various n-type dopants in Ge are presented in Fig. \ref{fig:Fig2}(b). We deduce from Fig. \ref{fig:Fig2}(a-b) that the n-type impurities can introduce both shallow and deep traps in Ge. These traps can impact the performance of Ge detectors by affecting their ability to detect traces of rare physical events in nature

\subsection{C. Defect complexes}
%%%%%%%%%%%%%%%%%%%%%%%%%%%%%%%%%%%%%%%%%%%%%%%%%%%%%%%%%%%%%%%%%%%%%%%
%%%%%%%%%%%%%%%%%%%%%%%%%%%%%%%%%%%%%%%%%%%%%%%%%%%%%%%%%%%%%%%%%%%%%%%
%%%%%%%%%%%%%%%%%%%%%%%%%%%%%%%%%%%%%%%%%%%%%%%%%%%%%%%%%%%%%%%%%%%%%%%
\subsubsection{Stability of defect complexes}
It has been found experimentally\cite{C18,C19}  and theoretically\cite{C20,C21, A53, A55, A56}  that the diffusion of n-type dopants in Ge is facilitated by Ge vacancies. A neutral Ge vacancy has a high formation energy of 3.82 eV, meaning it can only be introduced under non-equilibrium growth conditions such as ion implantation or electron irradiation.\cite{C42, C43, C44, C45}  Once formed, Ge vacancies can interact with n-type dopants to form defect complexes. To assess their stability, we analyze the binding energies of these defect complexes. The binding energy ($\rm E_b$) of a defect complex can be expressed mathematically as:\cite{A47,A48}

\begin{equation}
\rm {E_b} = \Delta{E_{f}^{complex}} - {\sum \Delta{E_{f}^{isolated}}}
 \label{E4}
\end{equation}
where, $\rm {\Delta E_f}^{complex}$ and $\rm {\Delta E_f}^{isolated}$ are formation energies of a defect complex and isolated defect. The summation in Equation \ref{E4} is over all the defects that form a complex. A positive $E_b$ indicates repulsion between defects, thus such defect complex is not stable. In contrast, a negative $\rm E_b$  indicates that the defect complex is stable. 
%%%%%%%%%%%%%%%%%%%%%%%%%%%%%%%%%%%%%%%%%%%%%%%%%%%%%%%%%%%%%%%%%%%%%%%
\begin{figure}[ht!]
\centering
\includegraphics[width=12cm]{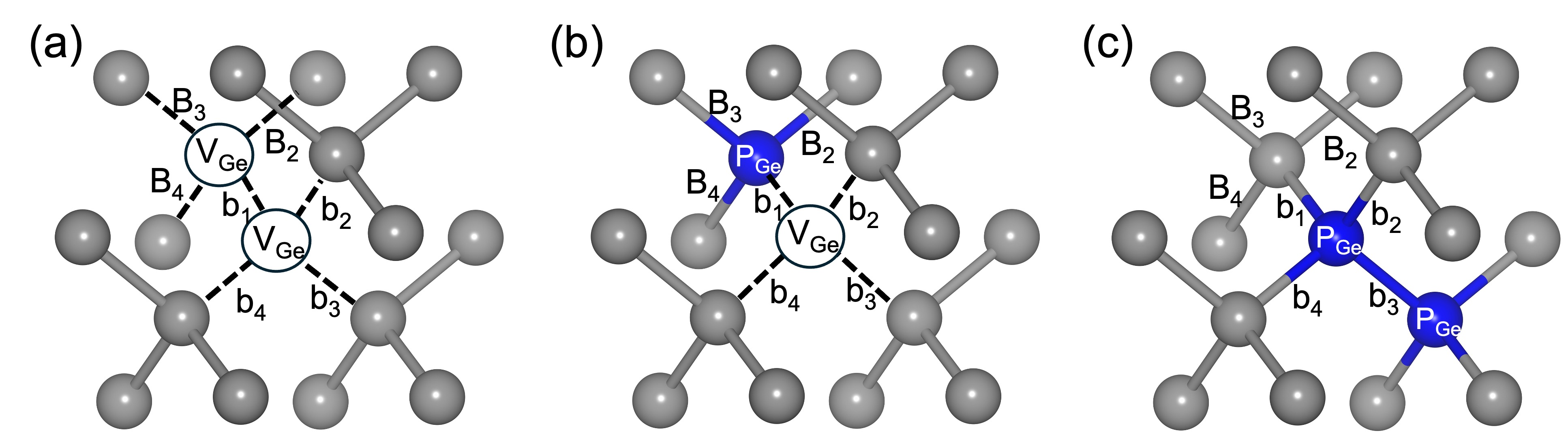}
\caption{Illustration of defect complexes considered in this study for complexes with two defects. (a) VV complex, (b) PV complex, and (c) PP complex. The gray and blue spheres represent Ge and Phosphorus (P) atoms, respectively. }
\label{fig:Fig3}
\end{figure}
%%%%%%%%%%%%%%%%%%%%%%%%%%%%%%%%%%%%%%%%%%%%%%%%%%%%%%%%%%%%%%%%%%%%%%%
\begin{figure}[ht!]
\centering
\includegraphics[width=8cm]{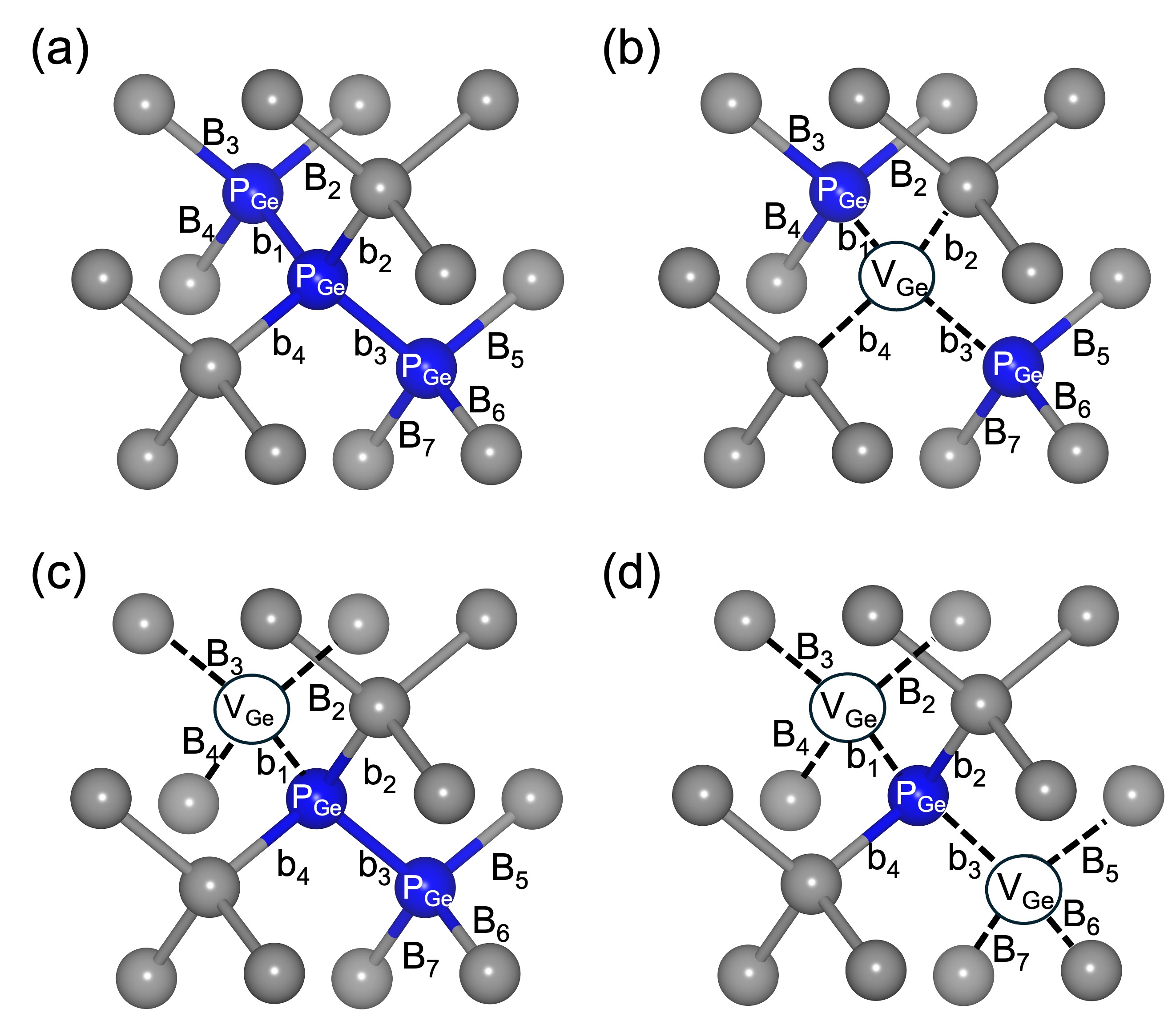}
\caption{Illustration of defect complexes considered in this study for complexes with three defects. (a) PPP complex, (b) PVP complex, (c) PPV complex, and (d) VPV-complex. The gray and blue spheres represent Ge and P atoms, respectively. }
\label{fig:Fig4}
\end{figure}
%%%%%%%%%%%%%%%%%%%%%%%%%%%%%%%%%%%%%%%%%%%%%%%%%%%%%%%%%%%%%%%%%%%%%%%
In this study, we investigate the defect complexes formed from pairs and triplets of isolated defects, as illustrated in Fig. \ref{fig:Fig3} and  Fig. \ref{fig:Fig4}. 
For convenience, we denote the $\rm V_{Ge}V_{Ge}$ complex simply as VV and $\rm P_{Ge}P_{Ge}$ complex as PP.  Similarly, $\rm P_{Ge}P_{Ge}P_{Ge}$ , $\rm P_{Ge}V_{Ge}P_{Ge}$, $\rm P_{Ge}P_{Ge}V_{Ge}$ and $\rm V_{Ge}P_{Ge}V_{Ge}$ complexes are denoted as PPP, PVP, PPV, and VPV, respectively. Defect complexes involving N, As, and Sb are represented using the same notation.

We begin by examining the stability  of a neutral VV complex in Ge. The calculated formation energy of a neutral VV complex in this study is 6.03 eV. Our calculations yield a binding energy ($\rm E_b$) of  -1.61 eV for the VV complex, indicating strong binding between the two vacancies upon formation.  This finding is consistent with previous experimental and theoretical studies, which suggest that vacancies in Ge have a tendency to cluster and form voids.\cite{C46,C18,C19, C20,C21, A53, A55, A56} 

%%%%%%%%%%%%%%%%%%%%%%%%%%%%%%%%%%%%%%%%%%%%%%%%%%%%%%%%%%%%%%%%%%%%%%%%%%%%%%%%%%%%%%%%%%%%%%%%%%%%%%%%%%%%%%%%%%%%%%%%%%%%%%%%%%%%%%%%%%%%%%%%%%%%%%%%%%%%%%%%%%%%%%%%%%%%%%%%%%%%%%%%%%%%%%%%%%%%%%%%%%%%%%%%%%%%%%%%%%%%%%%%%%%%%%%%%%%%%%%%%%%%%%%%%%%%%%%%%%%%%%%%%%%%%%%%%%%%%%%%%%%%
%%%%%%%%%%%%%%%%%%%%%%%%%%%%%%%%%%%%%%%%%%%%%%%%%%%%%%%%%%%%%%%%%%%%%%%
\begin{table}[ht!]
\caption{\label{tab:table4} Calculated formation ($\rm \Delta E_f$) and binding ($\rm E_b$) energies of neutral defect complexes at HSE06 level. D represents N, P, As, and Sb dopant.}
\centering
%\begin{tabular}{ m{15mm} m{15mm}  m{15mm} m{15mm} m{15mm} m{15mm} } 
\begin{tabular}{c c c c c c c c}
\hline\hline
Dopant    &                 & DD         & DV     & VDV     & DDD     & DVD    & DDV\\
\hline
$\rm N $ & $\rm \Delta E_f$ (eV) & \rm 3.28   & 4.69  & \rm 8.25    & 5.69    & 5.11    & 6.99 \\
         & $\rm E_b$ (eV)        & \rm -2.26  & -1.90 & \rm -2.16   & -2.62   & -4.25   & -2.37 \\
 \hline
$\rm P $ & $\rm \Delta E_f$ (eV) & \rm 1.05   & 3.08  & \rm 6.84    & 1.65   & 2.16      & 3.10 \\
         & $\rm E_b$ (eV)        & \rm -0.03  & -1.28 & \rm -1.35   & 0.02   & -2.74     & -1.81 \\
 \hline
 $\rm As $ & $\rm \Delta E_f$ (eV)  & \rm 1.38   & 3.06  & \rm 6.69   & 1.92   & 2.18    & - \\
         & $\rm E_b$ (eV)        & \rm  0.09   & -1.40  & \rm -1.60   & -0.01   & -2.94   & - \\
 \hline
 $\rm Sb $ & $\rm \Delta E_f$ (eV) & \rm 1.91   & 3.14  & \rm  6.39   & 3.25   & 2.43  & - \\
         & $\rm E_b$ (eV)        & \rm  0.26    &  -1.51 & \rm -2.07   & 0.78   & -3.05   & - \\
 \hline
\hline
\end{tabular}
\end{table}
%%%%%%%%%%%%%%%%%%%%%%%%%%%%%%%%%%%%%%%%%%%%%%%%%%%%%%%%%%%%%%%%%%%%%%%%%%%%%%%%%%%%%%%%%%%%%%%%%%%%%%%%%%%%%%%%%%%%%%%%%%%%%%%%%%%%%%%%%%%%%%%%%%%%%%%%%%%%%%%%%%%%%%%%%%%%%%%%%%%%%%%%%%%%%%%%%%%%%%%%%%%%%%%%%%%%%%%%%%%%%%%%%%%%%%%%%%%%%%%%%%%%%%%%%%%%%%%%%%%%%%%%%%%%%%%%%%%%%%%%%%%%%%%%%%%%%%%%%%%%%%%%%%%%%%%%%%%%%%%%%%%%%%%%%%%%%%%%%%%%%%%%%%%%%%%%%%%

Table \ref{tab:table4} presents the computed formation and binding energies for neutral defect complexes in Ge. 
Our calculations indicate that, unlike NN and NNN complexes, which exhibit strong binding between N dopants, 
DD and DDD complexes (where D = P, As, and Sb) are not stable in Ge. This instability is reflected in their positive or nearly positive binding energies, as shown in Table \ref{tab:table4}. 
We further predict that vacancies can form stable defect clusters with n-type dopants, as evidenced by their negative binding energies in Table \ref{tab:table4}. This result aligns with previous experimental and theoretical studies that also suggest the formation of defect clusters in Ge.\cite{C18, C19, C20,C21, C46, A53, A55, A56} 
Additionally, our calculations reveal that N forms the most strongly bound defect complexes, although the formation energies of N-related complexes with vacancies are larger than formation energies of P, As, and Sb related defect complexes. 
Among the studied defect structures, DVD complexes (where D = N, P, As, and Sb) exhibit significantly stronger binding compared to DDV and VDV complexes. This suggests that n-type dopants preferentially occupy the nearest neighboring sites on either side of the vacancy when forming defect complexes, a behavior that has also been observed experimentally in a previous study \cite{C18}.
We also infer from our calculations that larger defect complexes with vacancies exhibit significantly stronger binding than the smaller ones. For instance, the binding energy for a PVP complex is -2.74 eV, compared to the -1.28 eV for a PV complex. We also find that vacancy in Ge lower  its formation energy by forming defect complexes with P, As, and Sb. For example, the formation energy of a PVP ($\sim$2.16 eV) complex is lower than a PV ($\sim$3.08 eV) complex, which in turn is lower than the formation energy an isolated Ge vacancy ($\sim$3.82 eV).  This reduction in formation energy upon clustering, along with the strong binding of these defect complexes, suggests that larger complexes are energetically more favorable in Ge.   Our calculations also reveal that AsAsV and SbSbV complexes relax to the respective AsVAs and SbSbV complexes, respectively, indicating a preference for these rearranged structures. The computed binding energies for defect complexes are in good qualitative agreement with previous studies conducted at GGA level.\cite{A52, A54} However, in this work, we examined the stability of these complexes using a more accurate hybrid functional approach.

Similar to Ge vacancies and other n-type defects, the formation of defect complexes in Ge alters its local structural properties. The structural modifications, specifically in terms of Ge-defect and defect-defect bond lengths for the simplest defect complexes, are summarized in Table S1. Our calculations indicate that structural changes near the defect site are more pronounced for N-related defect complexes compared to those involving P, As, and Sb. This difference arises from variations in atomic size and electronegativity among the n-type dopants studied here. However, beyond the immediate vicinity of the defect complex, no significant structural changes are observed.

\subsubsection{Formation Energy and charge transition level for defect complexes}

\begin{figure}[ht!]
\centering
\includegraphics[width=14cm]{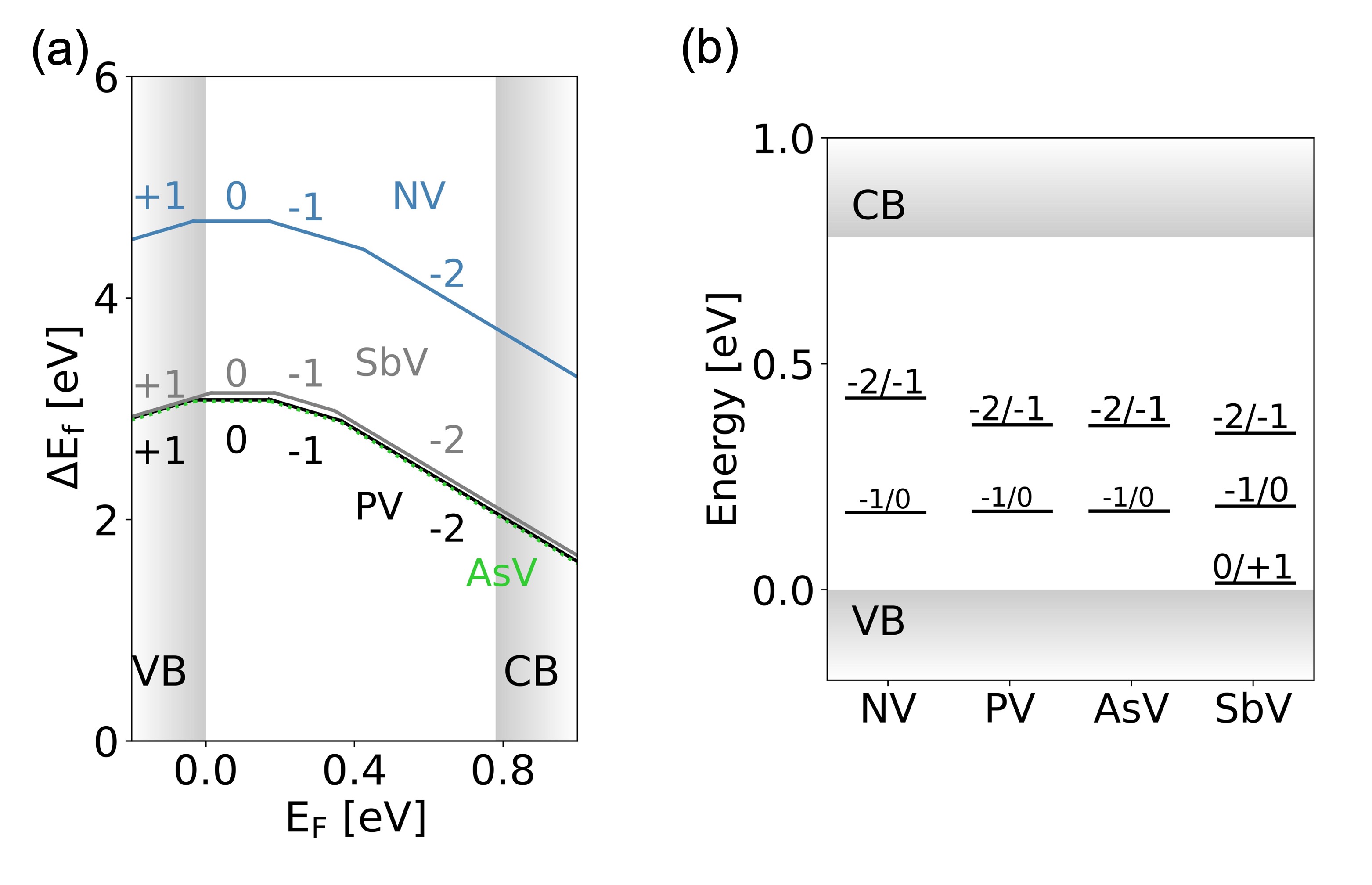}
\caption{(a) Formation energy ($\rm \Delta E_f$) as a function of the Fermi level ($\rm E_F$) for defect complexes (b) Charge transition levels for defect complexes. The shaded region in plots represents the valence band (VB) and conduction band (CB).}
\label{fig:Fig5}
\end{figure}

Next, we examine the energetics of the simplest defect complexes formed by the clustering of a single vacancy with a single n-type dopant in Ge. Fig. \ref{fig:Fig5} presents the formation energy as a function of electron chemical potential for these defect clusters. Our calculations reveal that, similar to \(\rm V_{Ge}\), defect complexes primarily exist in the 0, -1, and -2 charge states within the bandgap of Ge. The formation energy of an NV complex in these charge states is significantly higher than that of PV, AsV, and SbV complexes. For instance, the formation energy of a neutral NV complex is approximately 4.69 eV, whereas the formation energy for PV, AsV, and SbV complexes is around 3.10 eV. Due to their similar formation energies across different charge states, PV, AsV, and SbV complexes are expected to have comparable concentrations in Ge.  
Furthermore, Fig. \ref{fig:Fig5} indicates that defect-vacancy complexes are more likely to be in a doubly negative charge state under n-type doping conditions, consistent with a previous theoretical study performed using the GGA+U method \cite{C20}.

Fig. \ref{fig:Fig5}(a-b) illustrates that these defect complexes exhibit -1/0 and -2/-1 charge transition levels within the electronic bandgap of Ge. Similar to the formation energies shown in Fig. \ref{fig:Fig5}, our calculations indicate that PV, AsV, and SbV complexes have nearly identical charge transition levels. The -1/0 and -2/-1 transitions for these complexes are located at approximately 0.17 eV and 0.35 eV, respectively. This suggests that their charge transition levels are very close to the corresponding ionization levels of \(\rm V_{Ge}\). Notably, the SbV complex exhibits an additional 0/+1 transition level near the valence band edge at 0.015 eV. However, this defect level is relatively shallow and less detrimental compared to the deeper traps associated with SbV cluster. Unlike PV, AsV, and SbV complexes, the -2/-1 ionization level for the NV complex is slightly deeper, located at 0.42 eV. However, its -1/0 ionization level (~0.17 eV) is similar to other defect complexes considered in this study.  
Fig. \ref{fig:Fig5} also confirms that, like \(\rm V_{Ge}\), these defect complexes behave as multi-level acceptors, despite being composed of donor-type dopants. Table \ref{tab:table5} summarizes the computed charge transition levels for these defect complexes and compares them with previously reported theoretical values. The discrepancies between the transition levels in the two studies are likely due to differences in supercell size and the computational method used in Ref. \citenum{C20}.

%%%%%%%%%%%%%%%%%%%%%%%%%%%%%%%%%%%%%%%%%%%%%%%%%%%%%%%%%%%%%%%%%%%%%%%
\begin{table}[ht!]
\caption{\label{tab:table5} Calculated charge transition level (eV) for defect complexes. The values in the parenthesis are obtained using GGA+U method in Ref. \citenum{C20}.}
\centering
\begin{tabular}{c c c c c c  } % centered columns (4 columns)
\hline\hline
 Transitions & NV  &  PV & AsV & SbV  \\
\hline
0/+1 & - & - &  - & 0.015   \\
-1/0 & 0.17 & 0.17 (0.28)\textsuperscript{\emph{a}} &  0.17 (0.26)\textsuperscript{\emph{a}}  & 0.18 (0.17)\textsuperscript{\emph{a}}    \\
-2/-1 & 0.42 & 0.37 (0.52)\textsuperscript{\emph{a}} & 0.36 (0.47)\textsuperscript{\emph{a}} & 0.35 (0.18)\textsuperscript{\emph{a}}   \\ 
\hline\hline
\end{tabular}

  %\textsuperscript{\emph{b}} Ref. \citenum{C17};
  \textsuperscript{\emph{a}} Ref. \citenum{C20}
\end{table}
%%%%%%%%%%%%%%%%%%%%%%%%%%%%%%%%%%%%%%%%%%%%%%%%%%%%%%%%%%%%%%%%%%%%%%%

The presence of deep and shallow traps in Ge, arising from defects and their complexes with vacancies, as predicted in this study, are expected to affect the performance of HPGe detectors by creating charge trapping sites. Deep traps are particularly detrimental  because the trapped carriers lack sufficient energy to escape from deep states, leading to permanent charge loss. The carriers in shallow traps, on the other hand, can undergo continuous trapping and de-trapping phenomena due to thermal energy gain $\sim$$\rm k_B$T. In either case, the mobility of these carriers and the charge collection rates at the collecting electrodes are affected. This compromises the performance of HPGe detectors. Therefore, it is crucial to minimize or eliminate these defects during the synthesis process to ensure optimal detector performance.

\section{IV. Conclusions}

In summary, we employed a hybrid functional within density functional theory (DFT) to investigate possible n-type group-V defects and their complexes in Ge. Our predictions indicate that P, As, and Sb have lower formation energies compared to N dopants, Ge vacancies, and Ge interstitials. These defects (P, As, and Sb) are found to form shallow traps below the conduction band edge, approximately at 0.64 eV and 0.75 eV from the valence band maximum.
Additionally, n-type defects in Ge readily form defect complexes with Ge vacancies, exhibiting strong binding, with larger complexes binding more strongly than smaller ones. Furthermore, larger complexes of P, As, and Sb with vacancies have lower formation energies than smaller ones, suggesting a preference for forming defect clusters in Ge. For example, the formation energy of a PVP complex (~2.16 eV) is lower than that of a PV complex (~3.08 eV), which, in turn, is lower than the formation energy of a vacancy (~3.82 eV) in Ge. The impurity-vacancy complexes introduce deep traps in Ge, approximately at 0.17 eV and 0.35 eV, measured with respect to the valence band maximum. These traps are expected to cause charge trapping in HPGe detectors, affecting the measured waveforms in experiments and leading to energy resolution degradation. We believe our results will make a significant contribution to understanding possible n-type defects in Ge and aid in developing necessary mitigation strategies to eliminate them during the fabrication process of HPGe detectors, thereby improving their performance in measuring rare physical events such as  $\textit{0}\nu\beta\beta$.

\section{Associated Content}
\subsection{Supporting Information}

The Supporting Information is available free of charge on the XXX  at DOI: XXX

\section*{Author Information}

\textbf{Corresponding Authors}
\\
Gaoxue Wang*, Los Alamos National Laboratory, Los Alamos, New Mexico 87545, USA. {{\textbf{Email}}: gaoxuew@lanl.gov}
\\
Sandip Aryal*, Los Alamos National Laboratory, Los Alamos, New Mexico 87545, USA. {{\textbf{Email}}: saryal@lanl.gov}
\newline
\textbf{Author}
\\
Enrique R. Batista, Los Alamos National Laboratory, Los Alamos, New Mexico 87545, USA. {{\textbf{Email}}: erb@lanl.gov}

\section{Notes}
The authors declare no competing financial interest.

\section{Data Availability}
The data that support the findings of this study are available from the authors upon request.

\section{Acknowledgment}
The authors gratefully acknowledge the funding for this work from the Laboratory Directed Research and Development program at Los Alamos National Laboratory (LANL) under project 20230047DR. Los Alamos National Laboratory is operated by Triad National Security, LLC, for the National Nuclear Security Administration of the U.S. Department of Energy (contract no. 89233218CNA000001). This research used resources provided by the LANL Institutional Computing Program, which is supported by the U.S. Department of Energy National Nuclear Security Administration under Contract No. 89233218CNA000001. 
%%%%%%%%%%%%%%%%%%%%%%%%%%%%%%%%%%%%%%%%%%%%%%%%%%%%%%%%%%%%%%%%%%%%%
%%%%%%%%%%%%%%%%%%%%%%%%%%%%%%%%%%%%%%%%%%%%%%%%%%%%%%%%%%%%%%%%%%%%%
%%%%%%%%%%%%%%%%%%%%%%%%%%%%%%%%%%%%%%%%%%%%%%%%%%%%%%%%%%%%%%%%%%%%%
%%%%%%%%%%%%%%%%%%%%%%%%%%%%%%%%%%%%%%%%%%%%%%%%%%%%%%%%%%%%%%%%%%%%%
%%%%%%%%%%%%%%%%%%%%%%%%%%%%%%%%%%%%%%%%%%%%%%%%%%%%%%%%%%%%%%%%%%%%%
%%%%%%%%%%%%%%%%%%%%%%%%%%%%%%%%%%%%%%%%%%%%%%%%%%%%%%%%%%%%%%%%%%%%%
%%%%%%%%%%%%%%%%%%%%%%%%%%%%%%%%%%%%%%%%%%%%%%%%%%%%%%%%%%%%%%%%%%%%%
%%%%%%%%%%%%%%%%%%%%%%%%%%%%%%%%%%%%%%%%%%%%%%%%%%%%%%%%%%%%%%%%%%%%%
%%%%%%%%%%%%%%%%%%%%%%%%%%%%%%%%%%%%%%%%%%%%%%%%%%%%%%%%%%%%%%%%%%%%%
%%%%%%%%%%%%%%%%%%%%%%%%%%%%%%%%%%%%%%%%%%%%%%%%%%%%%%%%%%%%%%%%%%%%%
%%%%%%%%%%%%%%%%%%%%%%%%%%%%%%%%%%%%%%%%%%%%%%%%%%%%%%%%%%%%%%%%%%%%%
%%%%%%%%%%%%%%%%%%%%%%%%%%%%%%%%%%%%%%%%%%%%%%%%%%%%%%%%%%%%%%%%%%%%%
%%%%%%%%%%%%%%%%%%%%%%%%%%%%%%%%%%%%%%%%%%%%%%%%%%%%%%%%%%%%%%%%%%%%%
%%%%%%%%%%%%%%%%%%%%%%%%%%%%%%%%%%%%%%%%%%%%%%%%%%%%%%%%%%%%%%%%%%%%%
\bibliography{refs}
%%%%%%%%%%%%%%%%%%%%%%%%%%%%%%%%%%%%%%%%%%%%%%%%%%%%%%%%%%%%%%%%%%%%%
%%%%%%%%%%%%%%%%%%%%%%%%%%%%%%%%%%%%%%%%%%%%%%%%%%%%%%%%%%%%%%%%%%%%%
%%%%%%%%%%%%%%%%%%%%%%%%%%%%%%%%%%%%%%%%%%%%%%%%%%%%%%%%%%%%%%%%%%%%%
%%%%%%%%%%%%%%%%%%%%%%%%%%%%%%%%%%%%%%%%%%%%%%%%%%%%%%%%%%%%%%%%%%%%%
%%%%%%%%%%%%%%%%%%%%%%%%%%%%%%%%%%%%%%%%%%%%%%%%%%%%%%%%%%%%%%%%%%%%%
%%%%%%%%%%%%%%%%%%%%%%%%%%%%%%%%%%%%%%%%%%%%%%%%%%%%%%%%%%%%%%%%%%%%%

\end{document}